\documentstyle[sprocl]{article}

\input{psfig}

\def\PLB{{\em Phys. Lett.}  B}
\def\PRL{{\em Phys. Rev. Lett.}}
\def\PRD{{\em Phys. Rev.} D}

\def\be{\begin{equation}}
\def\ee{\end{equation}}
\def\bea{\begin{eqnarray}}
\def\eea{\end{eqnarray}}

\begin{document}

\title{EXCLUSIVE MESON PRODUCTION AT COMPASS} 
\author{Josef Pochodzalla$^a$, Lech Mankiewicz$^b$, Murray Moinester$^c$, Gunther Piller $^b$,\\ 
Andrzej Sandacz$^d$, Marc Vanderhaeghen$^e$} 

\address{
$^a$Max-Planck-Institut f\"ur Kernphysik, 69029 Heidelberg, Germany\\
$^b$Physics Department, Technical University Munich, D-85747 Garching, Germany\\ 
$^c$School of Physics and Astronomy, R. and B. Sackler Faculty of Exact Sciences, Tel 
Aviv University, 69978 Ramat Aviv, Israel\\
$^d$Soltan Institute for Nuclear Studies, 00681 Warsaw, Poland\\ $^e$University Mainz, 
D-55099 Mainz, Germany}


\maketitle\abstracts{ We explore the feasibility to study exclusive meson production 
(EMP) in hard muon-proton scattering $\mu ~p \rightarrow \mu ~p~M$ at the COMPASS 
experiment. These measurements constrain the off-forward parton distributions (OFPD's) of 
the proton, which are related to the quark orbital contribution to the proton spin.} 

\section{Introduction}
\label{lab_sec_1} The internal structure of the proton has been a subject of high 
interest for many decades. In general, the spin of the proton resides in its quark and 
gluon constituents and in their orbital angular momenta: 

\begin{equation}
\frac{1}{2}\langle \Delta q_v\rangle + \frac{1}{2}\langle \Delta S\rangle + \langle 
\Delta G\rangle + L_q +L_g = \frac{1}{2} 
\end{equation}

\noindent Here, $\frac{1}{2}\langle \Delta q_v\rangle$ denotes the spin in the valance 
quarks, $\frac{1}{2}\langle \Delta S\rangle$ is the total spin carried by the sea quarks 
and antiquarks, $\langle \Delta G\rangle$ represents the spin residing in the gluonic 
field, $L_q$ and $L_g$ represent the orbital momenta of the quarks and gluons, 
respectively. A main goal of the COMPASS experiment at CERN is the determination of the 
gluon contribution $\Delta G$. In COMPASS, $\Delta G$ will be studied via the 
photon-gluon fusion process, tagged either by charm-anticharm production, or by 
high-$p_T$ hadron pair production \cite{Bra98}. 

A factorization theorem for deeply-virtual Compton scattering (DVCS) was provided by Ji 
\cite{Ji97a} and Radyushkin \cite{Rad96,Mus97}. For exclusive meson production by 
longitudinally polarized photons, a factorization theorem was proven recently by Collins, 
Frankfurt, and Strikman \cite{Col97}. The theorem allows one to separate the meson 
production amplitudes into a perturbatively calculable component, describing the 
interaction of the virtual photon with quarks and gluons of the target, and a matrix 
element component which contains all information about the long-distance non-perturbative 
strong interaction dynamics in the produced meson and the nucleon target. The latter 
component can be parameterized, at leading order PQCD, in terms of 4 generalized 
structure functions, the so called {\em skewed} or {\em off-forward parton distributions} 
(OFPD's). Ji showed  \cite{Ji97a} that the second moment of these OFPD's determines the 
total quark angular momentum, the sum of the intrinsic quark spins and their orbital 
angular momentum. The OFPD's differ from the usual parton distributions probed in 
inclusive reactions by having a non-zero momentum transfer between the proton in the 
initial and final state. Therefore, the OFPD's are not accessible in standard inclusive 
measurements. They can, however, be measured in deeply-virtual Compton scattering or in 
hard exclusive lepto-production of mesons (EMP). Thus, exclusive meson production 
provides a unique tool to explore the contribution of orbital angular momentum and the 
valance quark polarization to the nucleon spin. 

EMP are important not only because they relate to the quark orbital momentum contribution 
to the proton spin. They should test our understanding of QCD generally, and of OFPD's in 
particular. COMPASS may lead in the study of these new processes, that are only now 
calculable in QCD.

\begin{figure}[t]

\begin{minipage}[t]{0.35\linewidth}
\psfig{figure=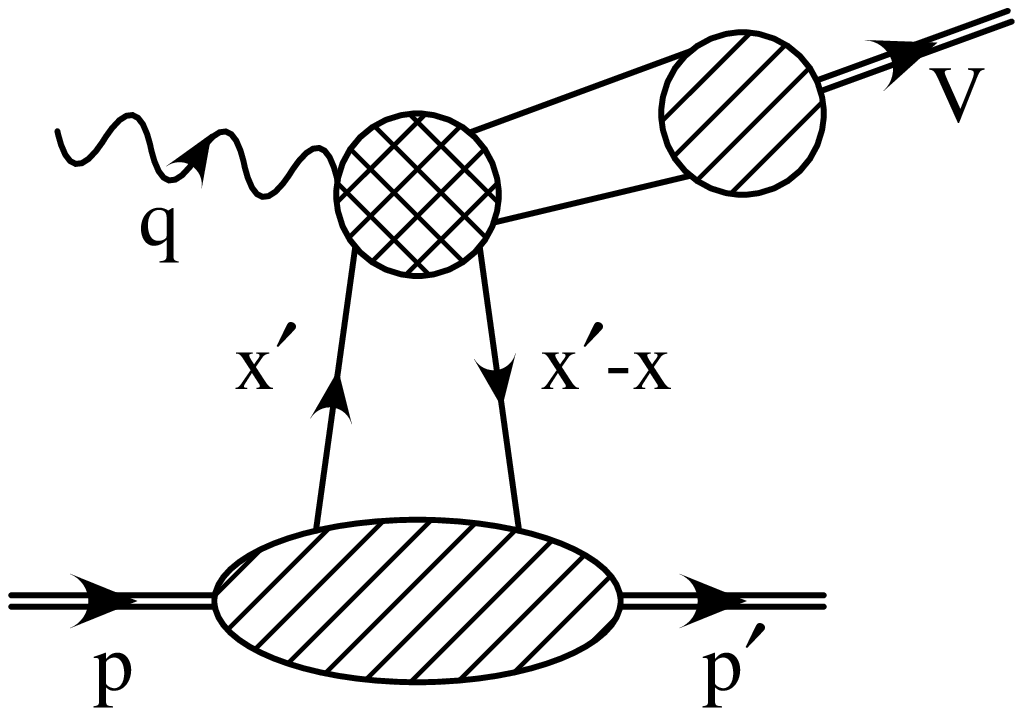,height=1.5in} \caption{Illustration of the 
factorization theorem for exclusive meson production \protect{\cite{Col97}}. 
\label{fig:fact}} 
\end{minipage}
\hspace{\fill} 
\begin{minipage}[t]{0.55\linewidth}
\psfig{figure=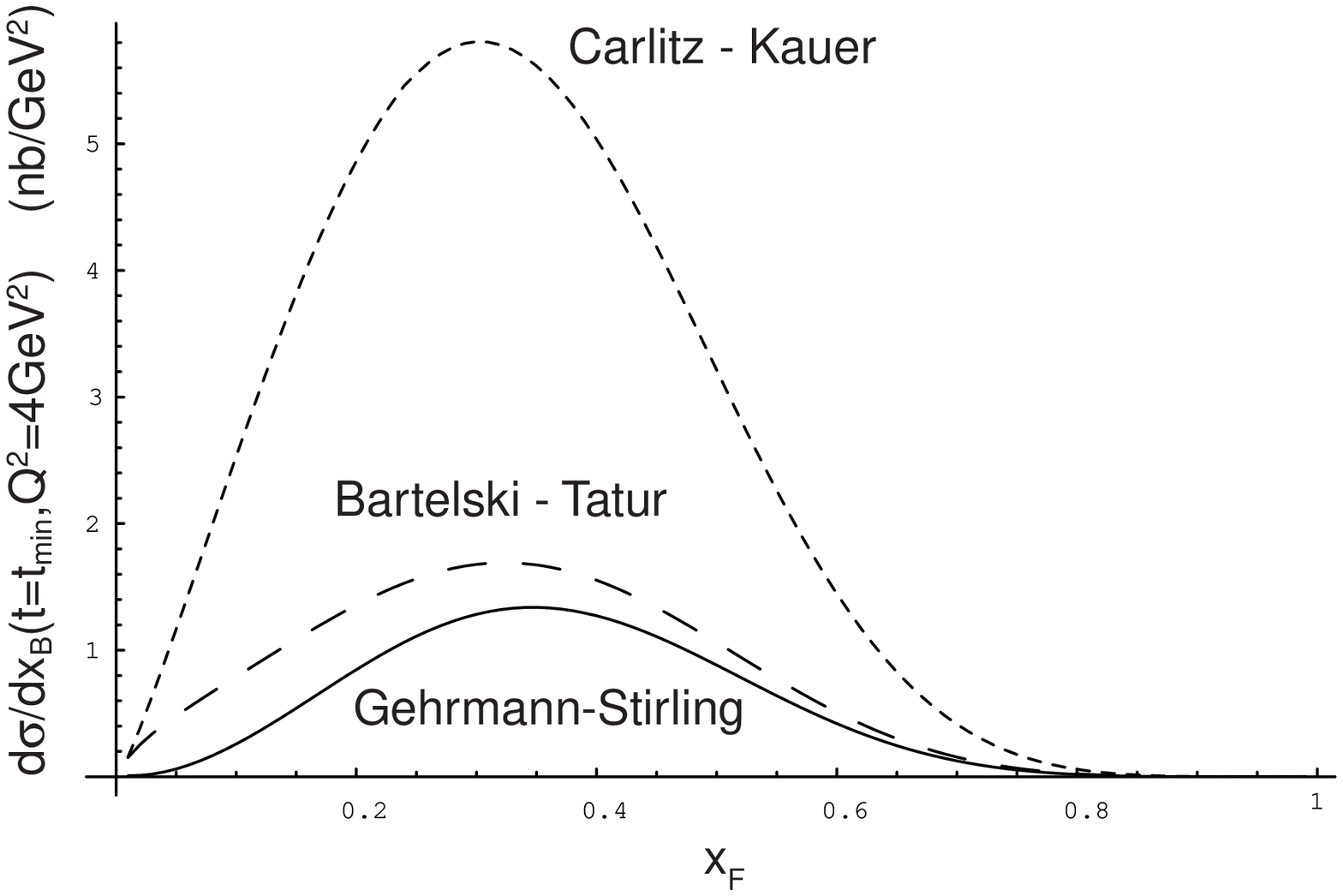,height=1.7in} \caption{Differential cross section for 
$\pi^0$ production at t = t$_{min}$, $Q^2 = 4 ~GeV^2$, as a function of x$_B$, in units 
of $nb/GeV^2$, for the three different valence quark parametrizations 
\protect{\cite{Geh96,Car77,Gos96,Kur98}}. \label{fig:dsigdt}} 
\end{minipage}
\end{figure}

In principle, Deeply Virtual Compton Scattering represents the most elementary process to 
access the OFPD's. This reaction has been considered for COMPASS \cite {dHo98}. It was, 
however, realized that this channel may suffer from potentially large backgrounds. 
Detailed understanding of (exclusive) meson production is therefore a precondition for 
studying DVCS. In turn, DVCS may represent an important complement for the Exclusive 
Meson Production study, which we discuss below. 

With the high muon beam energy and high luminosity in COMPASS \cite{compass} experiment, 
large $Q^2$ values can be reached with reasonable rates. For $\rho_0$ for example, a 
$Q^2$ range between 2 and 20 (80) GeV$^2$ is accessible for incident 100 GeV/c (200 
GeV/c) muon beam. Thus, this COMPASS program extends the approved activity of the CLAS 
collaboration at JLAB \cite{Did98}, which will explore the OFPD in a limited $Q^2$ range 
up to about 4 GeV$^2$. We distinguish between EMP channels in which the final state meson 
($\rho^0$, $\pi^+$ \cite{Man98}, $\phi$, etc.) can be detected via magnetic 
spectrometers, and those which require electromagnetic calorimetry. The charged cases can 
be detected with higher resolution and more easily without EM calorimetry. 

As another important aspect of this study, we also consider EMP of a "di-pion" (both 
resonant and non-resonant), such as $\pi^+\pi^-$, $\bar K K$, $\pi K$, as studied 
recently by Polyakov \cite{MVP}. These works demonstrated that the shape (not absolute 
cross sections!) of these "di-pion" mass distributions carry important information about 
quark distribution in the pion (kaon), and also about the structure of resonances ($\rho, 
f_2, K^*$, etc.). 

\section{The Proposed Experiment}
\label{lab_sec_3} 

\subsection{Experimental methodology and trigger}

We base our considerations on the scattered muon trigger that is forseen in the COMPASS 
experiment to study spin dependent DIS. The trigger fires if the muon is scattered at 
large angle $\theta_{\mu}$. In our simulations we assumed a good angular acceptance for 
the range $ 10(5) 
< \theta _{\mu } < 50 \: \rm{mrad} \: $ for 100 (200) GeV beam energy. 
The trigger is based on the X-Y correlations between hodoscope elements at about 31 and 
50 meters downstream of the target. 

Average recoil proton energies for EMP are quite low ($\sim$30-40 MeV). Thus, the recoil 
proton kinematics for COMPASS do not allow installing an efficient proton recoil detector 
to isolate "elastic" reactions. In order to be able to make optimal use of the 
experimental setup employed by the main muon program, or even to run initially in 
parallel with the main program, we will evaluate the inelasticity $I \propto 
(M_x^2-M_p^2)$ using measured momenta of the incident and final muons and produced meson 
\cite{San98}. The resolution will, however, not be sufficient to cleanly resolve events 
with a recoil proton or a recoil $\Delta^+$ or N$^*$ in the final state. However, the 
OFPD's for these processes are related \cite {CEBAF98}. Thus, a comparison of the data 
should be made to model calculations which combine both, proton, $\Delta^+$ and N$^*$ 
production. 

The main goal of the proposed program is to obtain high statistics EMP data on a pure 
{\it unpolarized} proton target during dedicated data runs. This would also help to avoid 
"dilution" by the presence of different target nuclei in the polarized proton target. 
Also, with the liquid hydrogen target, the detection efficiency of produced $\pi^0$ decay 
$\gamma$'s would increase due to the bigger total radiation length of the LH. However, in 
order to explore the feasibility of this program already during the initial data taking 
period of COMPASS, we assumed the standard polarized target for our present simulations 
and count rate estimates. 

\subsection{Cross section for exclusive meson production (EMP)}

We estimate the dependence of the cross section for the reaction $\mu ~p \rightarrow \mu 
~p ~\pi^0$ on the kinematic variables $Q^2$, x$_B$, and t, using models for off-forward 
parton distributions. Following \cite{Rad96,Mus97}, we constructed the relevant 
off-forward parton distributions starting from polarized valence quark distributions. 

The exact QCD calculation involves time-consuming numerical integration, which is 
inconvenient for Monte Carlo purposes. However, neglecting QCD logarithmic evolution of 
off-forward parton distributions, one can obtain a simple parametrization based on a fit 
to the results of the exact formulae: 

\begin{equation}
\label{EQ_XSEC1} 
 \frac{d\sigma}{dt}(\gamma_L~ p \rightarrow \pi^0~ p) =
\frac{d\sigma}{dt}(x,Q^2)\bigg|_{t={\rm t_{min}}} \times e^{B(t-t_{min})}, 
\end{equation}

where: 

\begin{equation}
\label{EQ_XSEC2} \frac{d\, \sigma}{d\,t}(x,Q^2)\bigg|_{t={\rm t_{min}}} = 
  {{{{{ \alpha_S}^2({Q^2})}}\cdot{\rm PF}(x,{Q^2})\cdot
      {\rm UF}(x)}\over {{Q^2(Q^2+M^2)^2}}}.
\end{equation}

\noindent Here, $\alpha_S(Q^2)$ is the one-loop QCD coupling constant, which -- as usual 
for LO calculations -- was evaluated with $\Lambda_{QCD} = 0.2$ GeV and with $NF = 4$ for 
the number of active flavors. In this equation, ${\rm PF}(x,{Q^2})$ is given by: 
\begin{equation}
\label{EQ_XSEC3} {\rm PF}(x,Q^2) = 
  {{{{{Q^4}}}\,\left( 1 - x \right) }\over
    {{{\left( -{\it Q^2} - 0.881721\,x + {\it Q^2}\,x \right) }^2}}}.
\end{equation}
\noindent ${\rm UF}(x)$ is the part of the cross-section which is independent of $Q^2$. 
It would acquire a logarithmic $Q^2$ dependence when QCD evolution effects are included. 
The shape of ${\rm UF}(x)$ depends on the model of the off-forward parton distributions 
used in the actual calculation. 

Figure \ref{fig:dsigdt} shows the differential cross section for $\pi^0$ production at t 
= t$_{min}$, $Q^2 = 4~ GeV^2$, as a function of Bjorken x$_B$, in units of $nb/GeV^2$, 
for the three different valence quark parametrizations. The solid line refers to the 
Gehrmann-Stirling polarized valence parametrization \cite{Geh96}, the short-dashed line 
is based on the Carlitz-Kaur model \cite{Vdh97,Car77,Gos96} and the long-dashed line 
results from the Bartelski-Tatur polarized valence distribution \cite{Kur98}. We do not 
include twist-2 contributions to the OFPD, which according to recent calculations 
\cite{PPPPBGW97,PW99} may change the absolute cross sections 
for EMP. 

\subsection{Monte Carlo results}

The events of exclusive $\pi ^0$ production, $\mu + p \rightarrow \mu  + p + \pi ^0$ with 
the subsequent decay $\pi ^0 \rightarrow \gamma \gamma $, were generated with rates 
proportional to the estimates of differential cross sections from Ref. 18. 
The directly measured cross section could be expressed in terms of the virtual 
photoproduction cross section: 
\begin{equation}
\label{EQ_MC1} 
 \frac{{\rm d}\sigma _{\mu N \rightarrow \mu N \pi ^0}(\nu,
Q^2, t)} {{\rm d}\nu {\rm d}Q^2 {\rm d}t} = \Gamma (Q^2,\nu ) \frac{{\rm d}\sigma 
_{\gamma ^*N \rightarrow N \pi ^0}(x=Q^2/(2 M_p \nu), Q^2, t)}{{\rm d}t}, 
\end{equation}
where $\Gamma$ is the flux of virtual photons: 
\begin{equation}
\label{EQ_MC2} \Gamma = \frac{\alpha (\nu -\frac{Q^2}{2M_p})}{2\pi Q^2 E^2_{\mu } 
(1-\epsilon)}\: . 
\end{equation}
Here, $\alpha $ is the fine structure constant, and $\epsilon $ is the virtual photon 
polarization.

The cross sections were estimated for proton and carbon targets. The latter one is more 
representative for the complex material of the COMPASS polarized target. For the proton 
target, ${\rm d}\sigma _{\gamma ^*N \rightarrow N \pi ^0}(x, Q^2, t)/{\rm d}t$ is given 
by the Eqs. (\ref{EQ_XSEC1}-\ref{EQ_XSEC3}) and $B=4.5$ GeV$^{-2}$. For $x=0.1$, $Q^2=4 
\: \rm{GeV}^2$ and $~t=t_{min}$, $\frac{d\sigma}{dt}$ is equal to 0.88 nb/GeV$^2$. For 
similar conditions, Vanderhaeghen {\it et al.} \cite {Vdh97} give about 4 times larger 
cross section. We will therefore give counting rate estimates in this report with an 
uncertainty, which reflects the difference between these two theoretical estimates. The 
present theoretical uncertainties are mainly due to different assumptions on the 
shapes/magnitudes of the valence quark spin distributions. For the carbon target, the 
cross section was assumed to be the product of the proton cross section and a function: 
\begin{equation}
\label{EQ_MC4} 
 f(t) = 8.12 \: e^{ -47.7 (t-t_{min})} + 0.3 \: .
\end{equation}
This function was obtained from the parametrization of the NMC data for exclusive $\rho 
^0$ production on a carbon target \cite{NMC}. 

In our simulations, we considered the kinematic range of $\label{EQ_MC5} 1~\rm{GeV}^2 < 
Q^2 < Q_{max}^2\: \rm{GeV}^2 \: , $ and $ \label{EQ_MC6} 10 (15) \rm{GeV}< \nu < 85 
(180)\: \rm{GeV}$. Here, $Q_{max}^2 = 4 E_{\mu } (E_{\mu }-\nu )\cdot\sin 
^2(\theta^{max}_{\mu}/2)$, $\theta^{max}_{\mu} = 50 \: \rm{mrad}$ , and $E_{\mu }$ is the 
incident muon energy. The values outside (inside) the brackets correspond to the beam 
energy of 100 (200) GeV. The same convention will be used hereafter. The upper cut on 
$\nu $ was chosen to eliminate the kinematic region where the amount of radiative events 
is large, whereas the lower one to eliminate events with poor kinematic resolution. The 
value of $Q_{max}^2$ is about 20 (90) GeV$^2$. The cross section integrated within this 
kinematic regime 

\begin{figure}[t]

\hspace{2cm} \psfig{figure=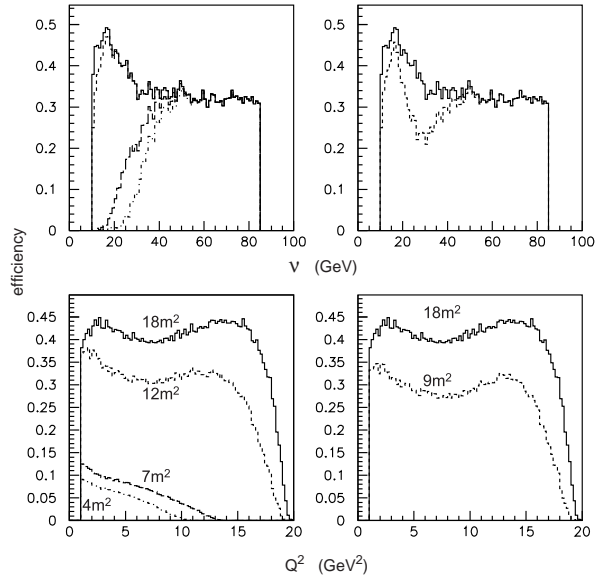,height=3.0in} 

\caption{The total acceptance to observe exclusive $\pi ^0$ events as a function of $\nu 
$ (top) and $Q^2$ (bottom) at a beam energy of 100 GeV. The different histograms 
correspond to different geometrical acceptances of the electro-magnetic calorimeter 
\label{fig:mc3}} 
\end{figure}

The total acceptance $\epsilon_{\pi ^0}$ to observe an exclusive $\pi ^0$ event, taking 
into account the geometric acceptance, the absorption of the decay photons before 
entering the calorimeter, and the reconstruction of $\pi ^0$ from the calorimetric 
measurements, is shown in the top parts of Fig. \ref{fig:mc3} as function of $\nu $ by 
the solid histograms. The main effect on $\epsilon_{\pi ^0}$ is due to a poor 
reconstruction of low energy $\pi ^0$ mesons from the calorimetric measurements, at small 
$\nu $, and due to the absorption of photons in the polarized target, most significantly 
at large $\nu $. The average total efficiency $\epsilon_{\pi ^0}$ is 0.43 for both 
incident beam energies. For the kinematic range $\nu  > 15$ GeV, $Q^2 > 1 \: {\rm 
GeV}^2$, $0.02 < x 
< 0.2$, the efficiency is quite large, varying between 0.3 and 0.5. 

Taking the above values of cross sections, and the expected luminosity for the COMPASS 
experiment \cite{compass} of $43 \; \rm{pb}^{-1} \rm{day}^{-1}$, we evaluate the number 
of deep inelastic exclusive $\pi ^0$ events per day, $N_{\mu \pi ^0 N}^{tot}$/day$(Q^2 > 
1\: \rm{GeV}^2)$, to be about 248-992 (53-212) for 100 (200) GeV. After taking into 
account effects of acceptance, secondary interactions, and the efficiency of $\pi ^0$ 
reconstruction, the number of events, $N_{\mu \pi ^0 N}^{cuts}$/day$(Q^2 
> 1\: \rm{GeV}^2)$, is about 107-428 (23-92). The expected numbers  of exclusive $\pi ^0$ events 
for a running period of 150 days (corresponding to a 1 year period) assuming  an overall 
(SPS and COMPASS) efficiency of 25\% are 4013-16050 (855-3420). For $Q^2 > 4 GeV^2$, we 
expect only 130 (42) counts. Adopting the Carlitz-Kauer valence quark parametrization the 
last numbers will increase by about a factor of 4. Thus, in 1 year of COMPASS operation 
we will collect at most 500 $\pi^0$ events with $Q^2 > 4 GeV^2$. 

Up to this point we assumed full size electro-magnetic calorimeters (total of 18~m$^2$) 
at their default positions. The different curves in Fig. \ref{fig:mc3} show the result if 
only limited coverage is assumed. While a reasonable efficiency may still be obtained 
with about half the calorimeter size (9~m$^2$), a further cut (e.g. down to 7 m$^2$) 
significantly reduces the $\pi^0$ efficiency. Furthermore, the count rates presented 
above were based on a polarized target. Assuming that the overall length of a target is 
limited to typically 2 m, a dedicated LH target would lead to a reduced luminosity (about 
a factor of 4) unless the beam intensity can be correspondingly increased.   

The detailed discussion of the possibility for COMPASS experiment to measure another 
exclusive vector meson production channel, $\mu + N \rightarrow \mu  + N + \rho ^0$, is 
presented in Ref. \cite{San98}. Here we only briefly summarize the results of the Monte 
Carlo calculations for this reaction. The kinematic range is somewhat narrower than for 
the previous channel: 
$
2 GeV^2< Q^2 < Q_{max}^2$ and $20 (40) < \nu < 90 (180)\: \rm{GeV}$. The estimated value 
of the measured cross section 
$\sigma _{\rho 
^0}^{th}$, is equal to 0.46 nb (0.5 nb) for 100 GeV (200 GeV). These values are obtained 
using a parametrization of the cross section measured in the NMC experiment \cite{NMC}. 
For the given kinematic range, the average total acceptance $\epsilon_{\rho ^0}$ is 0.73 
(0.68). Taking into account the expected luminosity for the COMPASS experiment, and the 
effects of acceptance and secondary interactions, the number of events with the invariant 
mass of two pions in the central part of $\rho ^0$ peak, $0.62 
< m_{\pi \pi } < 0.92 \; \rm{GeV}$,$N_{\mu \rho ^0 N}^{cuts}$/day$(Q^2 > 2~\rm{GeV}^2)$, 
is about 10$^4$ for both beam momenta. For a running period of 150 days assuming again an 
overall (SPS and COMPASS) efficiency of 25\% this will result in about 3.7$\cdot$10$^5$ 
events. For $Q^2 >~4 GeV^2$ we expect 75 k (107 k) events. 

\section{Conclusions}

We explored the feasibility to study exclusive meson production in hard muon-proton 
scattering $\mu ~p \rightarrow \mu ~p~M$ at the COMPASS experiment. For moderate $Q^2 
> 1~GeV^2$, a significant counting rate for exclusive $\pi^0$ production can be reached.
For the experimental conditions assumed in our calculations, still up to 500 events 
beyond a $Q^2$ of 4~GeV$^2$ can be collected during 1 year of operation.  

On the other hand, COMPASS may provide already during the first main $\mu$ runs, high 
statistics data on exclusive $\rho^0$ production even at $Q^2$ beyond values presently 
accessible at JLAB. Furthermore, data taken with the polarized target may also allow a 
first view at polarization phenomena. 
 
\section*{Acknowledgements}

This work was partially supported by the Polish State Committee for Scientific Research 
(KBN), by the U.S.-Israel Binational Science Foundation (BSF) and the Israel Science 
Foundation founded by the Israel Academy of Sciences and Humanities. 

------------------------------------------------------------------------

\end{document}